\documentclass[9pt,conference]{IEEEtran}
\IEEEoverridecommandlockouts
\usepackage{cite}
\usepackage{amsmath,amssymb,amsfonts}
\usepackage{algorithmic}
\usepackage{graphicx}
\usepackage{textcomp}
\usepackage{multirow}
\usepackage{xcolor}
\usepackage{filecontents}
\def\BibTeX{{\rm B\kern-.05em{\sc i\kern-.025em b}\kern-.08em
    T\kern-.1667em\lower.7ex\hbox{E}\kern-.125emX}}
\begin{document}

\title{GraFPrint: A GNN-Based Approach for Audio Identification\\

\thanks{\textsuperscript{*}A. Bhattacharjee and S. Singh contributed equally to this work.
A. Bhattacharjee and S. Singh are research students at the UKRI Centre for Doctoral Training in Artificial Intelligence and Music, supported jointly by UK Research and Innovation [grant number EP/S022694/1] and Queen Mary University of London.
}
}


\author{
\IEEEauthorblockN{Aditya Bhattacharjee, Shubhr Singh, Emmanouil Benetos}
\IEEEauthorblockA{
    School of Electronic Engineering and Computer Science, Queen Mary University of London, UK
}
}
\maketitle
\begin{abstract}
This paper introduces GraFPrint, an audio identification framework that leverages the structural learning capabilities of Graph Neural Networks (GNNs) to create robust audio fingerprints. Our method constructs a k-nearest neighbour (k-NN) graph from time-frequency representations and applies max-relative graph convolutions to encode local and global information. The network is trained using a self-supervised contrastive approach, which enhances resilience to ambient distortions by optimizing feature representation. GraFPrint demonstrates superior performance on large-scale datasets at various levels of granularity, proving to be both lightweight and scalable, making it suitable for real-world applications with extensive reference databases.
\end{abstract}
\section{Introduction}
\label{sec:intro}

Automatic audio identification is a process that matches a query audio snippet to a reference audio recording stored in a database, with the goal of accurately identifying the recording, even in noisy environments. This system encodes audio recordings into compact audio fingerprints, designed to be efficient for storage and retrieval while remaining robust against various acoustic distortions. In the context of identification of musical recordings, such a system has various applications such as identifying an unknown song in the presence of other noisy sources or for enforcing copyright in online content.

Over the past two decades, numerous approaches have been developed for automatic audio identification. Landmark-based methods \cite{wang2006shazam}\cite{six2014panako}\cite{sonnleitner2014quad} focus on generating audio fingerprints by extracting prominent peaks from time-frequency representations, such as spectrograms. To enhance the robustness of these peak-based fingerprints against transformations, binary hashing techniques are employed, mapping and storing the relative positions of the peaks in the spectrogram. In contrast, neural network-based approaches leverage self-supervised contrastive training~\cite{simclr} to learn an embedding space that is resilient to signal distortions. Architectures such as convolutional neural networks (CNNs)~\cite{chang2021neural} \cite{wu2022asymmetric} are commonly employed encoders, while more recent approaches have explored the use of self-attention and transformer models \cite{singh2022attention}\cite{singh2023simultaneously}. 

Unlike landmark-based methods, neural network architectures can learn latent, noise-invariant patterns directly from data, eliminating the need for feature engineering. The search and retrieval process in these systems typically employs efficient approximate nearest-neighbour search algorithms. However, overall system performance critically depends on the quality of the learned embeddings. As database sizes increase, particularly with short and noisy audio queries, retrieval accuracy can degrade if the embedding model lacks robustness. The key challenge lies in developing embedding models that maintain high discriminative power across various audio conditions and database scales. This highlights the need for advanced methods that can learn more robust and compact audio representations, effectively balancing accuracy and scalability while leveraging efficient search algorithms for reliable identification in real-world scenarios with extensive reference databases.

\begin{figure}[t]
    \centering
    \includegraphics[width=0.49\textwidth]{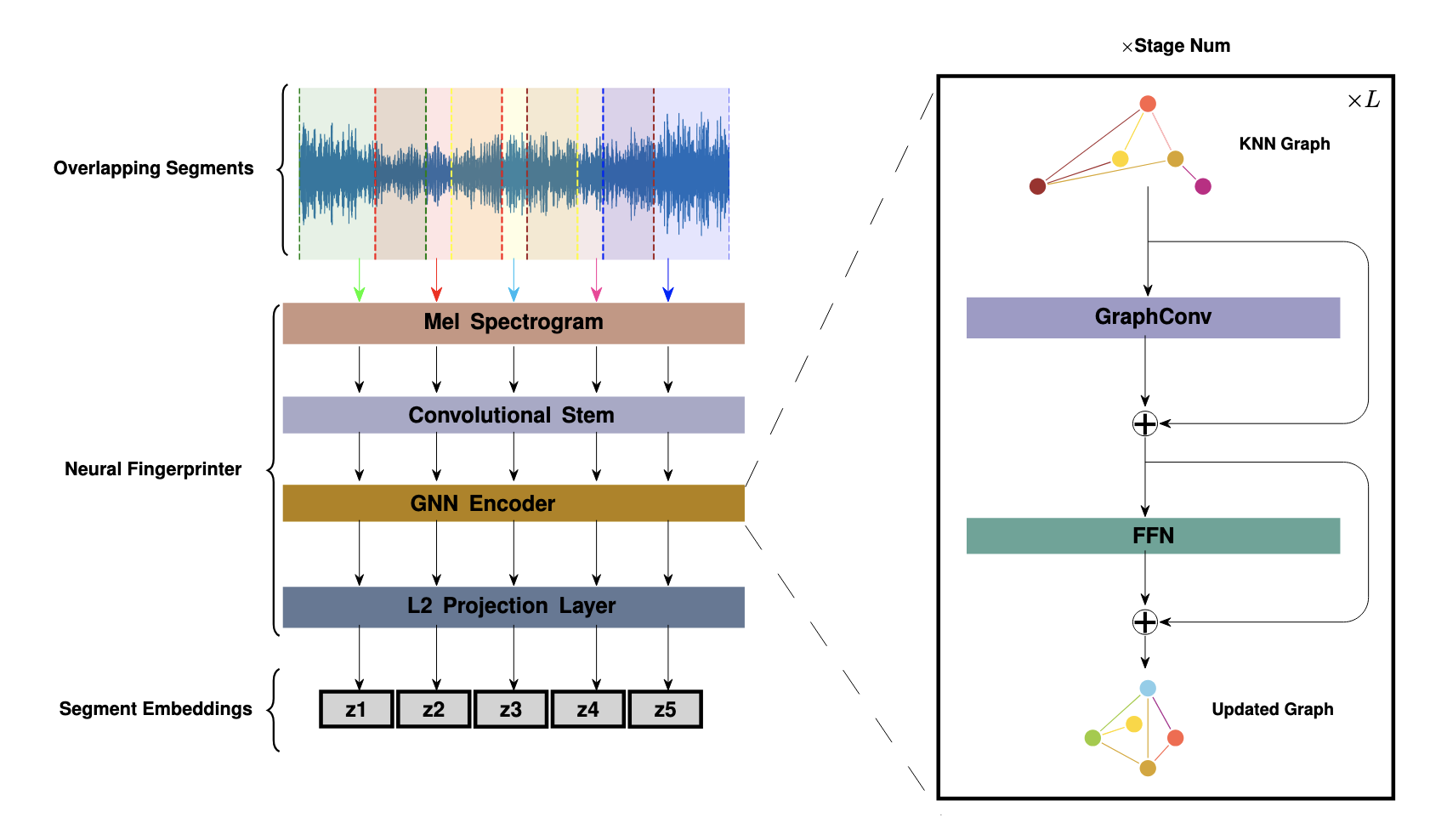}
    \caption{Overview of the \textit{GraFPrint} framework; overlapping audio segments are input into the contrastively trained neural fingerprinter. The output segment embedding space facilitates the approximate nearest neighbour search for matching query audio to the corresponding reference audio fingerprint.}
    \label{fig:pipeline}
\end{figure}

GNNs have recently gained attention for their ability to capture and preserve intricate structural patterns within data, even when the data is not naturally graph-structured. Unlike traditional methods, such as Convolutional Neural Networks (CNNs), which excel at processing grid-like data, GNNs are particularly effective in learning from non-Euclidean spaces by aggregating features across nodes based on their relationships, thus retaining important structural information \cite{li2019deepgcns}. In computer vision, GCNs have been adapted and benchmarked for image classification tasks \cite{han2022vision}. A similar approach \cite{singh2024atgnn} has been applied to audio tagging, where it has outperformed traditional CNN and attention-based architectures. 

Building on these successes, we propose the \textit{GraFPrint} (Graph-based Audio Fingerprint) framework. Inspired by the landmark-based audio fingerprinting approaches, which represent a spectrogram as a ``constellation map'' of time-frequency points, \textit{GraFPrint} models the latent relationships between points using graph neural networks. This approach harnesses the strengths of GNNs to improve the robustness and efficiency of audio identification. The main contributions of this work are:
\begin{itemize}
    \item A novel GNN-based approach for accurate and scalable automatic audio identification.
    \item Evaluation of our approach with a large reference database, demonstrating the efficacy of our lightweight GNN encoder in audio identification tasks.
    \item Benchmarking of our framework for matching queries to reference data at different levels of granularity, showcasing its adaptability for various use cases.
\end{itemize}
Our code has been made available for reproducibility at \texttt{https://github.com/chymaera96/GraFP}.

\section{GraFPrint Framework}
\label{sec:design}
In this section, we define the design of the framework. Figure \ref{fig:pipeline} shows the schematics of the architecture and the data pipeline. 

\subsection{Feature Extraction}
The input features for the neural network encoder are computed by randomly extracting a $t$-second-long audio segment $s$ from the input data. A log-mel spectrogram is computed from $s$. Let $S$ represent the original log-mel spectrogram with dimensions
with $F$ frequency bins and $T$ time frames. $S$ and its augmented view $S'$ (refer to section \ref{section:augmentation}) are used to form positive data pairs for the contrastive learning. In order to preserve the positional information of the time-frequency points, the spectrograms are concatenated with positional encodings for the frequency and time axis. For a batch $B$ of spectrograms, the resultant tensor $S \in \mathbb{R}^{B \times 3 \times F \times T}$ comprises three channels: a time index and a frequency index channel in addition to the spectrogram amplitudes.

\subsection{Encoder Network}

The feature map $S$ is passed through a neural network block that consists of strided convolutional layers. This allows the network to combine relevant local features in the time-frequency points and results in a more compact representation $X_0 \in \mathbb{R}^{B \times C \times H \times W}$. Here, \( H \), \( W \), and \( C \) represent the height, width, and number of channels, respectively.

Let $P^{(c)}$ denote a single data example in $X_0$. 

\begin{equation}
    P^{(c)} = \begin{pmatrix}
    p_{11}^{(c)} & p_{12}^{(c)} & \cdots & p_{1W}^{(c)} \\
    p_{21}^{(c)} & p_{22}^{(c)} & \cdots & p_{2W}^{(c)} \\
    \vdots & \vdots & \ddots & \vdots \\
    p_{H1}^{(c)} & p_{H2}^{(c)} & \cdots & p_{HW}^{(c)}
    \end{pmatrix}, \quad c = 1, 2, \ldots, C
\end{equation}

Each time-frequency point $p_{f,t}$ is transformed into a $d$-dimensional latent embedding using a non-linear projection layer $f(\cdot)$:

\[ f: \mathbb{R}^{B \times C} \rightarrow \mathbb{R}^{B \times d} \]

The projection is implemented through a $1 \times 1$ convolution layer followed by batch normalisation and non-linear activation. We flatten this feature map to obtain a set of projected points $P' \in  \mathbb{R}^{d \times N}$ which can be considered as an unordered set of $N = H \cdot W $ nodes in an undirected graph $G$.

To establish the graph structure, a k-nearest neighbour (k-NN) algorithm is applied. For each node in the set $P'$, the algorithm identifies its $k$ closest neighbours based on Euclidean distance in the feature space. 
 The graph structure, thus formed, is an abstraction for learning invariance to transformations (refer to section \ref{section:augmentation}) that may be relevant to the audio identification task. 

The node embeddings of the graph are subsequently refined through graph convolution layers designed to facilitate information exchange between neighbouring nodes via a message-passing operation. Specifically, the update for a node $x_i$ utilises a max-relative graph convolution \cite{li2019can} as follows:
\begin{equation}
g(\cdot) = x'' = \text{concat}(x_i, \max(x_j - x_i) \mid j \in N(x_i))
\end{equation}
\begin{equation}
h(\cdot) = x' = x''W_{\text{update}},
\end{equation}
where $N(x_i)$ denotes the set of neighbors for node $x_i$, and $x'$ and $x''$ represent the node embeddings updated through various operations. We refer to the combination of these operations as \textit{GraphConv}. To enhance feature diversity of the feature space and mitigate the risk of oversmoothing \cite{li2018deeper}, a linear layer is applied to each node before and after the \textit{GraphConv} operation, accompanied by a non-linear activation function. The updated graph convolution operation for a node can thus be represented as:
\begin{equation}
y_i = \sigma(\text{GraphConv}(W_{\text{in}}x_i))W_{\text{out}} + x_i,
\end{equation}
where $y_i$ denotes the updated node embedding, $\sigma$ represents a non-linear activation function, and $W_{\text{in}}$ and $W_{\text{out}}$ are the weights of fully connected layers applied respectively before and after the \textit{GraphConv} operation. This framework enables the dynamic construction of the graph, which adapts to the features and relationships within the data as they evolve through the network. This adaptability enhances the graph's ability to capture complex patterns and dependencies among the node embeddings, enabling it to learn a latent space which is robust to environmental distortions.

Subsequently, a feed-forward network (FFN) is applied on the refined node embeddings. Originally introduced within the Transformer architecture, an FFN typically consists of two fully connected linear layers separated by a non-linear activation function, such as ReLU.
Each \textit{GraphConv} and \textit{FFN} block is followed by a strided 2-D convolution layer to downsample the feature space. The downsampling layer reduces the number of nodes that are input to the subsequent layers, thus decreasing the overall computational burden of the k-NN graph computation.

A graph embedding is obtained by average-pooling the individual node embeddings in the graph. The resulting latent embedding is projected into a 128-dimensional representation $z$ using a fully-connected projection layer.

\subsection{Contrastive Training}

The encoder network is trained using a simple contrastive setup similar to the one used in \cite{chang2021neural}. Given a log-mel spectrogram $S$ with $F$ frequency bins and $T$ time frames, a data augmentation function $A(\cdot)$ uses a series of transformations to produce an augmented view $S' = A(S) \in \mathbb{R}^{F \times T}$ (refer to section \ref{section:augmentation}). 

The training objective aligns representations of different views of the same sample (forming positive pairs), bringing them closer in the embedding space while pushing apart representations of different samples (forming negative pairs). This is achieved using the normalized temperature-scaled cross-entropy (NT-Xent) loss function \cite{simclr}. Given a batch of $N$ samples, each with an augmented view, resulting in $2N$ views, the loss for a positive pair $(z_i,z_j)$ is:
\begin{equation}
\ell(i, j) = -\log \frac{\exp(\text{sim}(z_i, z_j) / \tau)}{\sum_{k=1}^{2N} \mathbb{1}_{[k \neq i]} \exp(\text{sim}(z_i, z_k) / \tau)}
\end{equation}
where $\text{sim}(\cdot)$ is cosine similarity, $\tau$ is a temperature parameter, and $\mathbb{1}_{[k \neq i]}$ is an indicator function. The total loss for the batch is the average overall positive pairs:
\begin{equation}
\mathcal{L} = \frac{1}{2N} \sum_{k=1}^{N} \left[ \ell(2k-1, 2k) + \ell(2k, 2k-1) \right]
\end{equation}

\subsection{Search and Retrieval}

The audio identification framework is evaluated for the robustness of the retrieval process against noisy transformations applied to small query audio segments. Given the query set and the reference database (see section \ref{sec:dataset}), we divide the audio waveforms into \( t \)-second-long overlapping segments and compute the audio fingerprints for each segment using the embedding model. Let $\{q_i\}_{i=1}^m$ be the set of audio fingerprints derived from the $m$ overlapping query segments. For each \( q_i \), we perform an approximate nearest-neighbour (ANN) search in our reference database to identify the fingerprints that match the query. The product-quantised inverted file index search (IVF-PQ), implemented using FAISS \cite{johnson2019billion}, is employed as the ANN algorithm. Let \( \mathcal{I} \in \mathbb{Z}^{+^{m \times n}} \) represent the set of indices retrieved by the ANN search, where $n$ denotes the number of probes used in the search. To accurately align the query and reference sequences, we apply an offset compensation to the retrieved indices \( \mathcal{I} \) to adjust for predicting the starting segment index. Specifically,
\begin{equation}
    \mathcal{I}^*[i, j] = \mathcal{I}[i, j] - i
\end{equation}

We extract the set of unique indices $\mathcal{C}$ from $\mathcal{I}^*$ and determine the matched reference index $\hat{i}$ by finding the candidate sequence that maximises the sequence-level similarity score, computed using the inner product of the query fingerprints and the corresponding offset-adjusted reference fingerprints. This is given by
\begin{equation}
    \hat{i} = \arg\max_{k \in \mathcal{C}} \left( \frac{1}{m} \sum_{j=1}^{m} \langle q_j, r_{j+k} \rangle \right),
\end{equation}
where \( r_{j+k} \) represents the reference fingerprint at the aligned index.

\section{Experimental Setup}
\label{sec:implementation}

\subsection{Experiment Details}

The proposed model is trained using two data-parallelised NVIDIA A100 GPUs for 400 epochs. The contrastive learning utilises the Adam optimizer and the learning rate during training is adjusted using a cosine decay scheduler. Table \ref{tab:hyper} provides the hyperparameters used for training and evaluation. 

\begin{table}[h]
    \centering
    \caption{Experimental configurations for the proposed framework}
    \label{tab:hyper}
    \begin{tabular}{|l|l|}
        \hline
        \textbf{Parameter} & \textbf{Value} \\
        \hline
        Size of training dataset & 8000 \\
        Sampling rate & 16,000 Hz \\
        log-power Mel-spectrogram size $F \times T$ & $64\times 32$ \\
        Fingerprint \{window length, hop\} & \{1s, 0.1s\} \\
        Fingerprint dimension & 128 \\
        Temperature parameter $\tau$ & 0.05 \\
        Batch size $B$ & 256 \\
        Number of epochs & 400 \\
        \hline
        \textbf{IVF-PQ configuration} & \\
        \hline
        Number of centroids & 256 \\
        Codebase size & $2^{64}$  \\
        \hline
    \end{tabular}
\end{table}

\subsection{Dataset}
\label{sec:dataset}
We use the \textit{Free Music Archive} (FMA) \cite{fma_dataset} dataset for our experiments. 
\begin{itemize}
    \item \textbf{Training set}: The \texttt{fma-small} subset containing 8K 30-second-long examples is used for training the models.
    \item \textbf{Reference database}: We conduct our experiments by computing the reference fingerprints using the \texttt{fma-medium} subset containing 25K 30-second long examples. Further, we scale up the search experiments with the \texttt{fma-large} subset containing 106K segments. 
    \item \textbf{Query database}: We extract two sets of 2000 query segments from each of the reference databases. The queries are 1 to 5-second long audio segments which are transformed using a test subset of the background noise and room impulse responses discussed in the next subsection. We have made the query sets available for reproducibility. 
\end{itemize}

\begin{table*}[!ht]
\centering
\scriptsize  
\caption{Top-1 hit rate(\%) comparison in the segment-level search in different noisy environments for varied query lengths. The reference database used for these experiments is derived from \texttt{fma-medium}.}
\label{tab:metrics}
\begin{tabular}{|c|c|c|c|c|c|c|c|c|c|c|c|}
\hline
\multirow{2}{*}{Method} & \multirow{2}{*}{Query Length} & \multicolumn{5}{c|}{Noise} & \multicolumn{5}{c|}{Noise + Reverb}  \\ \cline{3-12}
&  & 0dB & 5dB & 10dB & 15dB &20dB & 0dB & 5dB & 10dB & 15dB & 20dB  \\ \hline
\multirow{4}{*}{NAFP \cite{chang2021neural}} & 1s & 50.7 & 69.7 & 73.7 & 76.0 & 76.9 & 20.8 & 43.9 & 55.5 & 58.5 & 58.9 \\
& 2s & 71.0 & 83.4 & 85.5 & 87.6 & 87.5 & 37.8 & 65.6 & 73.5 & 75.2 & 76.7  \\
& 3s & 77.7 & 84.8 & 88.9 & 89.2 & 89.1 & 50.1 & 72.6 & 80.0 & 79.5 & 80.1  \\ 
& 5s & 82.6 & 89.2 & 90.2 & 90.5 & 91.2 & 60.2 & 79.1 & 83.4 & 82.8 & 83.1  \\ \hline

\multirow{4}{*}{TE + LSH \cite{singh2023simultaneously}} & 1s & \textbf{66.6} & 82.6 & 87.6 & 90.0 & 90.6 & 44.8 & 62.6 & 73.8 & 79.4 & 82.0  \\
& 2s & 80.4 & 88.2 & 91.6 & 93.2 & 94.8 & 63.4 & 78.2 & 84.6 & 86.0 & 85.4  \\
& 3s & 83.2 & 88.4 & 92.6 & 94.4 & 95.2 & 71.6 & 82.6 & 85.6 & 86.2 & 87.4  \\ 
& 5s & 85.6 & 90.0 & 92.8 & 94.2 & 95.8 & 80.0 & 87.0 & 87.1 & 87.6 & 87.2  \\ \hline

\multirow{4}{*}{TE + HT \cite{singh2023simultaneously}} & 1s & 64.9 & 81.7 & 87.9 & 90.1 & 90.9 & 41.9 & 62.0 & 74.6 & 79.2 & 81.2  \\
& 2s & 80.2 & 89.4 & 93.1 & 94.1 & 94.9 & 61.4 & 80.8 & 85.0 & 87.5 & 88.3  \\
& 3s & 84.7 & 90.8 & 95.5 & 96.3 & 97.1 & 70.7 & 84.1 & 88.6 & 89.0 & 90.1  \\ 
& 5s & 88.0 & 93.4 & 95.3 & 96.2 & 97.4 & 80.6 & 88.6 & 90.8 & 91.3 & 91.5  \\ \hline

\multirow{4}{*}{GraFPrint (Ours) } & 1s & 63.9 & \textbf{83.6} & \textbf{93.5} & \textbf{97.0} & \textbf{98.8} & \textbf{52.3} & \textbf{68.1} & \textbf{79.2} & \textbf{85.3} & \textbf{89.4}  \\
& 2s & \textbf{85.7} & \textbf{95.1} & \textbf{98.6} & \textbf{99.4} & \textbf{99.8} & \textbf{80.0} & \textbf{88.6} & \textbf{94.7} & \textbf{94.6} & \textbf{96.4}  \\
& 3s & \textbf{93.3} & \textbf{98.3} & \textbf{99.3} & \textbf{99.6} & \textbf{99.9} & \textbf{88.7} & \textbf{92.9} & \textbf{96.4} & \textbf{96.4} & \textbf{96.6}  \\ 
& 5s & \textbf{97.7} & \textbf{99.7} & \textbf{99.5} & \textbf{99.8} & \textbf{99.9} & \textbf{93.3} & \textbf{95.9} & \textbf{97.3} & \textbf{97.6} & \textbf{97.7}  \\ \hline

\end{tabular}
\end{table*}

\subsection{Data Augmentation}
\label{section:augmentation}

The following data augmentation methods are used in the contrastive training:
\begin{itemize}
    \item Time offset: to learn invariance to small time shifts that may occur in a query. The primary experiments are conducted with an offset of $\pm50$ms. Further, we investigate the effect of coarser audio identification use cases with larger time offsets (refer to Section \ref{sec:results}).
    \item Background noise mixing: simulates the presence of environmental noise in a real-life use case of audio identification by additive mixing of noise waveforms and the reference audio at different signal-to-noise ratios.  We use nearly 6 hours of noise recordings from the MUSAN dataset \cite{snyder2015musan}, which consists of ambient settings such as \textit{restaurant}, \textit{home} and \textit{street}.
    \item Convolutional reverb: ambient reverberation is simulated by convolving room impulse response (RIR) filters on the input waveform. We use the Aachen Room Impulse Response Database \cite{jeub2009binaural} for training and evaluation.
\end{itemize}

\subsection{Baseline Method}

As an effective benchmark, we implement a baseline inspired by the transformer-based neural encoder proposed in \cite{singh2023simultaneously}, which has shown promise for audio fingerprinting. In the absence of available implementations of
this approach, our implementation adapts the audio spectrogram transformer (AST) \cite{gong2021ast} architecture, incorporating the temporal patch embedding layer from \cite{singh2023simultaneously}. We train this baseline using a simple contrastive learning setup, closely following the input features and hyperparameters described in their work. 

\subsection{Evaluation Metrics}

The proposed framework is evaluated based on the top-1 hit rate on the query and reference databases. The top-1 hit rate for the audio identification task is the percentage of times the framework correctly retrieves the exact match from a reference database when given a set of noisy audio queries. This is given by
\begin{equation}
    \text{Top-1 Hit Rate} = \left( \frac{\text{Number of Correct Matches}}{\text{Total Number of Queries}} \right) \times 100\%
\end{equation}
A correct match is observed when the retrieved reference item matches the query within a time error margin. This margin determines the granularity of the audio identification process. Table \ref{tab:metrics} and \ref{tab:scaled} compares the performance of various frameworks within the allowed margin of $\pm50$ms. Further, we analyse the effect of changing the granularity on the retrieval rates.

\section{Results and Discussion}
\label{sec:results}

We test the robustness of our audio identification framework in comparison to reported metrics of other state-of-the-art frameworks. To compare the top-1 hit on our query set, we compute the audio fingerprints the reference database from \texttt{fma-medium}. Noisy queries are produced using the augmentation strategies discussed in section \ref{section:augmentation}. We benchmark performance under various signal-to-noise (SNR). To study the effect of impulse response convolution on the performance, we compare the metrics in the presence and absence of convolutional reverb. Table \ref{tab:metrics} shows that the \textit{GraFPrint} model consistently outperforms CNN and transformer-based setups. Retrieval rates are lower for smaller queries, as longer queries benefit from overlapping segments, making identification more reliable by mitigating single erroneous matches. We also observe that the presence of convolutional reverb leads to increased mismatches in all the benchmarked methods. For 1-second queries, our model shows an 11.6pp performance drop when reverb is introduced, compared to 21.8pp for the best baseline. Convolutional reverb is correlated to the original audio, leading to a more challenging augmentation scenario. 

To scale up our search experiments, we use a reference database derived from \texttt{fma-large}, which increases the possibility of mismatches during retrieval. Table \ref{tab:scaled} compares the retrieval rates of \textit{GraFPrint} with the AST baseline for queries with background mixing at different SNRs and in the presence of convolutional reverb. Under similar training and testing conditions, our framework outperforms the baseline by a minimum of 20.5pp across all conditions. Despite being trained with limited data, the \textit{GraFPrint} model generalizes effectively to a large reference database, demonstrating its robustness and efficiency. The scalability of \textit{GraFPrint} is further evidenced by its efficient use of computational resources. The encoder network has approximately 18M learnable parameters, compared to 45M in the AST baseline. Unlike the transformer encoder in \cite{singh2023simultaneously}, which requires large training batches, our setup showed only marginal performance gains with larger batches.
\begin{table}[h]
\centering
\scriptsize
\caption{Top-1 hit rate performance(\%) comparison for the segment-level search on a scaled reference database derived from \texttt{fma-large}.}
\begin{tabular}{|c|c|c|c|c|c|c|}
\hline
\multirow{2}{*}{Query Length} & \multirow{2}{*}{Method} & \multicolumn{5}{c|}{Noise + Reverb}  \\ \cline{3-7}
&  & 0dB & 5dB & 10dB & 15dB & 20dB  \\ \hline
\multirow{2}{*}{1s} & AST + IVFPQ & 22.2 & 26.7 & 29.3 & 37.0 & 40.8  \\
& GraFPrint &  \textbf{42.7} & \textbf{61.8} & \textbf{71.6} & \textbf{81.3}& \textbf{83.8}  \\ \hline

\end{tabular}
\label{tab:scaled}
\end{table}
\vspace{0pt}

\begin{figure}
    \centering
    \includegraphics[width=1\linewidth]{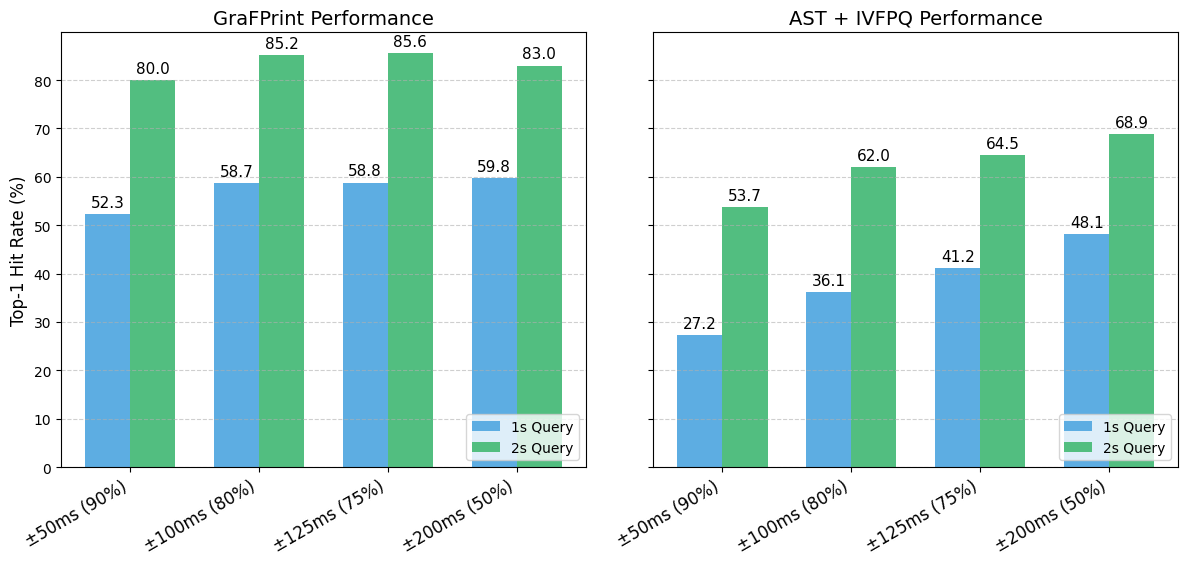}
    \caption{Comparison of top-1 hit rates at different levels of granularity. The horizontal axis shows the time offset used during training and the corresponding fingerprinting overlap percentage.}
    \label{fig:ablation}
\end{figure}

While finer alignment of the query and reference fingerprints can improve accuracy, it may reduce retrieval efficiency and robustness. To assess the impact of granularity on retrieval performance, we relax the time error margin during evaluation from $\pm50$ms to $\pm$\{100, 125, 250\}ms, and train models with corresponding time offsets in the augmentation process. This relaxation also enables the use of coarser fingerprints with less overlap. Figure \ref{fig:ablation} shows the top-1 hit rates vary across models trained with different time offsets consistent with four levels of granularity. For 2-second queries, the fine-grained approach leads to more overlapping fingerprints to offset possible mismatches. However, this is balanced out by the effect of a larger reference database. In contrast, 1-second queries, with only a single fingerprint, show lower retrieval rates due to a higher likelihood of mismatches. As the overlap decreases, so does the gap in retrieval rates between 1-second and 2-second queries ($ 27.7$pp$ \rightarrow13.2$pp for \textit{GraFPrint}). In addition to the effect of retrieval performance, a larger reference database has larger storage requirements. For instance, changing the overlap from 0.5s to 0.9s increases the reference database size by 5 times. 

\section{Conclusions}
\label{sec:conclusions}

In this work, we proposed a GNN-based embedding framework for audio identification that transforms time-frequency points into graph nodes, connected by their nearest neighbours in the feature space. This graph structure captures complex audio patterns, with node embeddings refined through graph convolutions and a feed-forward network to enhance robustness and discriminative power. The framework is trained using a self-supervised contrastive approach designed to learn invariance to the presence of ambient noise and reverberation. We observe that the model effectively handles ambient noise and reverberation, achieving competitive performance on large-scale databases and supporting both coarse and fine-grained alignment. A limitation of our graph-based approach is the slowdown in training due to the computational complexity of dynamically constructing and updating the k-NN graph, which worsens with larger datasets and more nodes. Exploring more efficient graph construction could mitigate this issue and enhance training efficiency. As our approach focuses on the neural architecture for learning the optimal and robust embedding space for identification of audio segments, we use simple quantization techniques for embedding storage and retrieval. This presents a potential for using the graph structure for data-driven hashing methods.


\bibliographystyle{IEEEtran}
\bibliography{refs}

\begin{thebibliography}{10}
\providecommand{\url}[1]{#1}
\csname url@samestyle\endcsname
\providecommand{\newblock}{\relax}
\providecommand{\bibinfo}[2]{#2}
\providecommand{\BIBentrySTDinterwordspacing}{\spaceskip=0pt\relax}
\providecommand{\BIBentryALTinterwordstretchfactor}{4}
\providecommand{\BIBentryALTinterwordspacing}{\spaceskip=\fontdimen2\font plus
\BIBentryALTinterwordstretchfactor\fontdimen3\font minus
  \fontdimen4\font\relax}
\providecommand{\BIBforeignlanguage}[2]{{%
\expandafter\ifx\csname l@#1\endcsname\relax
\typeout{** WARNING: IEEEtran.bst: No hyphenation pattern has been}%
\typeout{** loaded for the language `#1'. Using the pattern for}%
\typeout{** the default language instead.}%
\else
\language=\csname l@#1\endcsname
\fi
#2}}
\providecommand{\BIBdecl}{\relax}
\BIBdecl

\bibitem{wang2006shazam}
A.~Wang, ``The shazam music recognition service,'' \emph{Communications of the
  ACM}, vol.~49, no.~8, pp. 44--48, 2006.

\bibitem{six2014panako}
J.~Six and M.~Leman, ``Panako: a scalable acoustic fingerprinting system
  handling time-scale and pitch modification,'' in \emph{15th International
  Society for Music Information Retrieval Conference (ISMIR-2014)}, 2014.

\bibitem{sonnleitner2014quad}
R.~Sonnleitner and G.~Widmer, ``Quad-based audio fingerprinting robust to time
  and frequency scaling.'' in \emph{DAFx}, 2014, pp. 173--180.

\bibitem{simclr}
T.~Chen, S.~Kornblith, M.~Norouzi, and G.~Hinton, ``A simple framework for
  contrastive learning of visual representations,'' in \emph{International
  conference on machine learning}.\hskip 1em plus 0.5em minus 0.4em\relax PMLR,
  2020, pp. 1597--1607.

\bibitem{chang2021neural}
S.~Chang, D.~Lee, J.~Park, H.~Lim, K.~Lee, K.~Ko, and Y.~Han, ``Neural audio
  fingerprint for high-specific audio retrieval based on contrastive
  learning,'' in \emph{ICASSP 2021-2021 IEEE International Conference on
  Acoustics, Speech and Signal Processing (ICASSP)}.\hskip 1em plus 0.5em minus
  0.4em\relax IEEE, 2021, pp. 3025--3029.

\bibitem{wu2022asymmetric}
X.~Wu and H.~Wang, ``Asymmetric contrastive learning for audio
  fingerprinting,'' \emph{IEEE Signal Processing Letters}, vol.~29, pp.
  1873--1877, 2022.

\bibitem{singh2022attention}
A.~Singh, K.~Demuynck, and V.~Arora, ``Attention-based audio embeddings for
  query-by-example,'' \emph{arXiv preprint arXiv:2210.08624}, 2022.

\bibitem{singh2023simultaneously}
------, ``Simultaneously learning robust audio embeddings and balanced hash
  codes for query-by-example,'' in \emph{ICASSP 2023-2023 IEEE International
  Conference on Acoustics, Speech and Signal Processing (ICASSP)}.\hskip 1em
  plus 0.5em minus 0.4em\relax IEEE, 2023.

\bibitem{li2019deepgcns}
G.~Li, M.~Muller, A.~Thabet, and B.~Ghanem, ``Deepgcns: Can gcns go as deep as
  cnns?'' in \emph{Proceedings of the IEEE/CVF international conference on
  computer vision}, 2019, pp. 9267--9276.

\bibitem{han2022vision}
K.~Han, Y.~Wang, J.~Guo, Y.~Tang, and E.~Wu, ``Vision gnn: An image is worth
  graph of nodes,'' \emph{Advances in neural information processing systems},
  vol.~35, pp. 8291--8303, 2022.

\bibitem{singh2024atgnn}
S.~Singh, C.~J. Steinmetz, E.~Benetos, H.~Phan, and D.~Stowell, ``Atgnn: Audio
  tagging graph neural network,'' \emph{IEEE Signal Processing Letters}, 2024.

\bibitem{li2019can}
G.~Li, M.~M{\"u}ller, A.~Thabet, and B.~Ghanem, ``Can gcns go as deep as
  cnns,'' \emph{arXiv preprint arXiv:1904.03751}, pp. 1--17, 2019.

\bibitem{li2018deeper}
Q.~Li, Z.~Han, and X.-M. Wu, ``Deeper insights into graph convolutional
  networks for semi-supervised learning,'' in \emph{Proceedings of the AAAI
  conference on artificial intelligence}, vol.~32, 2018.

\bibitem{johnson2019billion}
J.~Johnson, M.~Douze, and H.~J{\'e}gou, ``Billion-scale similarity search with
  gpus,'' \emph{IEEE Transactions on Big Data}, vol.~7, no.~3, pp. 535--547,
  2019.

\bibitem{fma_dataset}
\BIBentryALTinterwordspacing
M.~Defferrard, K.~Benzi, P.~Vandergheynst, and X.~Bresson, ``{FMA}: A dataset
  for music analysis,'' in \emph{18th International Society for Music
  Information Retrieval Conference (ISMIR)}, 2017. [Online]. Available:
  \url{https://arxiv.org/abs/1612.01840}
\BIBentrySTDinterwordspacing

\bibitem{snyder2015musan}
D.~Snyder, G.~Chen, and D.~Povey, ``Musan: A music, speech, and noise corpus,''
  \emph{arXiv preprint arXiv:1510.08484}, 2015.

\bibitem{jeub2009binaural}
M.~Jeub, M.~Schafer, and P.~Vary, ``A binaural room impulse response database
  for the evaluation of dereverberation algorithms,'' in \emph{2009 16th
  International Conference on Digital Signal Processing}.\hskip 1em plus 0.5em
  minus 0.4em\relax IEEE, 2009, pp. 1--5.

\bibitem{gong2021ast}
Y.~Gong, Y.-A. Chung, and J.~Glass, ``Ast: Audio spectrogram transformer,''
  \emph{arXiv preprint arXiv:2104.01778}, 2021.

\end{thebibliography}

\end{document}